\documentclass[iop]{emulateapj}
\shorttitle{IC~3418} \shortauthors{Hester et al.}

\begin{document}
\title{IC~3418: Star Formation in a Turbulent Wake} 

\author{Janice A. Hester\altaffilmark{1}, Mark Seibert\altaffilmark{2}, James D. Neill\altaffilmark{1}, Ted K. Wyder\altaffilmark{1}, Armando Gil de Paz\altaffilmark{3}, Barry F. Madore\altaffilmark{2}, D. Christopher Martin\altaffilmark{1}, David Schiminovich\altaffilmark{4}, R. Michael Rich\altaffilmark{5}}
\email{jhester@srl.caltech.edu}

\altaffiltext{1}{California Institute of Technology, Pasadena, CA 91125, USA}
\altaffiltext{2}{Observatories of the Carnegie Institute of Washington, Pasadena, CA 91101, USA}
\altaffiltext{3}{Dpto. de Astrof\'{i}sica, Universidad Computense de Madrid, Madrid 28040, Spain}
\altaffiltext{4}{Department of Astronomy, Columbia University, New York, NY 10027, USA}
\altaffiltext{5}{Department of Physics and Astronomy, Division of Astronomy and Astrophysics, University of California, Los Angeles, CA 90095, USA}

\begin{abstract}
\textit{Galaxy Evolution Explorer} observations of IC~3418, a low surface brightness galaxy in the Virgo Cluster, revealed a striking 17~kpc UV tail of bright knots and diffuse emission. H$\alpha$ imaging confirms that star formation is ongoing in the tail. IC~3418 was likely recently ram pressure stripped on its first pass through Virgo.  We suggest that star formation is occurring in molecular clouds that formed in IC~3418's turbulent stripped wake.  Tides and ram pressure stripping (RPS) of molecular clouds are both disfavored as tail formation mechanisms. The tail is similar to the few other observed star-forming tails, all of which likely formed during RPS. The tails' morphologies reflect the forces present during their formation and can be used to test for dynamical coupling between molecular and diffuse gas, thereby probing the origin of the star forming molecular gas.
\end{abstract}


\section{Introduction}

\begin{figure}
\includegraphics[height=3in]{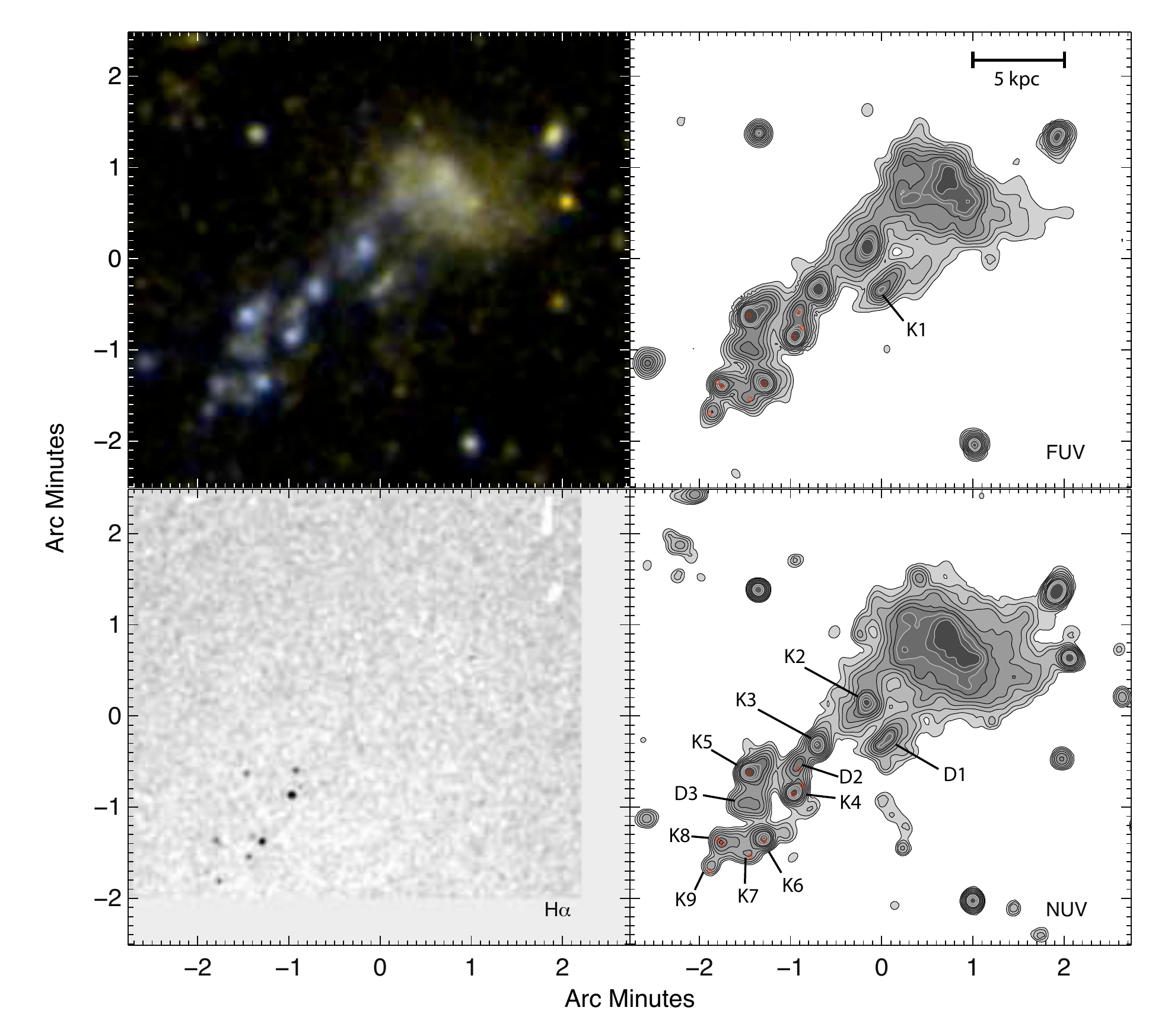}
\caption{Imaging showing a 17~kpc tail of star formation trailing IC~3418 as it plunges through Virgo's ICM. Upper left: color composite ultraviolet image; FUV is blue, NUV is red, and the average UV intensity is green.  Adaptive smoothing was performed using \textit{asmooth} from the \textit{XMM-Newton} Science Analysis Software package (http://xmm.esac.esa.int/sas). Upper right: FUV surface brightness. Contours are logarithmically spaced from 15e-17 to 200e-17 $\rm{erg~s^{-1}~cm^{-2}~arcsec^2}$.  Red circles indicate H$\alpha$ detections. The single knot defined in the FUV is labeled. Lower right: NUV surface brightness. Knots and diffuse regions discussed in the text are labeled. Lower left: \citet{Gavazzi06} continuum-subtracted H$\alpha$ image smoothed with an FWHM Gaussian of $0\arcsec.6$.}
\label{UV}
\end{figure}

IC~3418, a faint low surface brightness galaxy ($M_r=-16.6$) in the Virgo Cluster was observed with the \textit{Galaxy Evolution Explorer}~\citep[\textit{GALEX},][]{Martin05} and was found to have a remarkable ultraviolet tail of bright knots and diffuse emission (Figure~\ref{UV}).  The presence of near-ultraviolet (NUV), far-ultraviolet (FUV), and H$\alpha$~\citep{Gavazzi06} in the knots indicates that star formation is ongoing. Galaxies with similar star forming tails have been observed in other clusters at multiple wavelengths~\citep{Cortese07, Yoshida08, Sun06, Sun07}.  The UV morphology revealed in the moderately deep \textit{GALEX} images of IC~3418 is a potentially powerful probe of its tail's formation mechanism.

It has been suggested that similar tails are either star formation in ram pressure (RP) stripped molecular clouds~\citep{Yoshida08, Kronberger08, Kapferer09, Chung09} or a combination of cloud stripping and gravitational interactions~\citep[][C07]{Cortese07}. None of these tails appears like known tidal features, and they consist preferentially of gas and young stars.  We estimate that the tidal radii of all of these systems are beyond the disks' stellar radii. 

Star formation occurs in simulated RP stripped wakes when cold clouds and diffuse H~\textsc{i} are tightly dynamically coupled~\citep{Kronberger08, Kapferer09}.  Assuming strong dynamical coupling between the gas phases is likely invalid during ram pressure stripping (RPS). In simulations that resolve the multiphase medium, high density clouds are not stripped~\citep{Tonnesen09}.  For an observed example, isolated star forming clouds remain in the disk of NGC~4402, another Virgo spiral, despite stripping of the H~\textsc{i}~\citep{Crowl05}.  

We suggest that the tail of IC~3418 represents in situ molecular cloud and star formation in the galaxy's turbulent wake following RPS.  In Sections~\ref{ic3418} and~\ref{knots} we discuss the observations of IC~3418 and its star-forming tail. In Section~\ref{env} we show that while RPS of the H~\textsc{i} is expected, theory disfavors removing $\rm{H_2}$ from the galaxy via either stripping or tides.  In Section~\ref{morphology} we discuss the morphologies of IC~3418 and similar star forming tails; demonstrating that they favor in situ molecular cloud formation.

IC~3418 is beginning the transition between the blue and red sequences (${\rm NUV}-r=2.75$) and is thus interesting in its own right.   The stars in the tail may eventually contribute to the intracluster light (ICL), warranting a study of the frequency of these tails.  We limit our study to the tail's formation.

\section{IC~3418}
\label{ic3418}
\textit{GALEX} observed IC~3418 in the FUV (1350-1750~\AA) and the NUV (1750-2750~\AA) between 2004 March and 2006 May with effective exposure times of 2~ks and 15~ks, respectively.  An 1800~s H$\alpha$ observation was taken on the ESO 3.6~m in 2005 and is publicly available~\citep{Gavazzi03, Gavazzi06}. The galaxy was observed by the Sloan Digital Sky Survey~\citep[SDSS][]{York00}, but optical colors cannot be measured for the tail. We adopt a distance to Virgo of 16.5~Mpc~\citep{Mei07}, at which distance \textit{GALEX}'s resolution is 400~pc ($5\arcsec$).

IC~3418 is a member of the Virgo Cluster at a projected distance of 450~kpc ($94\arcmin$) from the cluster's core with a recessional velocity of $38~\rm{km~s^{-1}}$~\citep{Gavazzi04}.  Virgo is a dynamically young cluster, as evidenced by substructure in the galaxy distribution and the X-ray emission~\citep{Bingelli87, Schindler99}.  Galaxies with H~\textsc{i} tails which are presumably undergoing RPS are observed in Virgo's outskirts~\citep{Chung07}.  

\begin{figure}
\begin{center}
\includegraphics[height=2.5in]{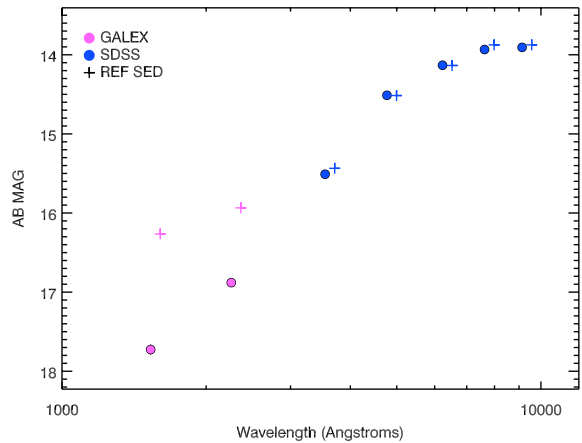}
\caption{UV (\textit{GALEX}) to optical (SDSS) SED of IC~3418's disk, measured within an elliptical aperture with a semimajor axis of $1\arcmin.2$, an axis ratio of 0.68, and a position angle of 67.8. Emission from the knots and foreground stars was masked. The reference SED represents the average of $\sim100$ \textit{GALEX}+SDSS galaxies with $-16.1 > M_r > -17.1$ and $0.3 < g-r < 0.45$.}\label{SED}
\end{center}
\end{figure}

\begin{figure}
\begin{center}
\includegraphics[height=2.25in]{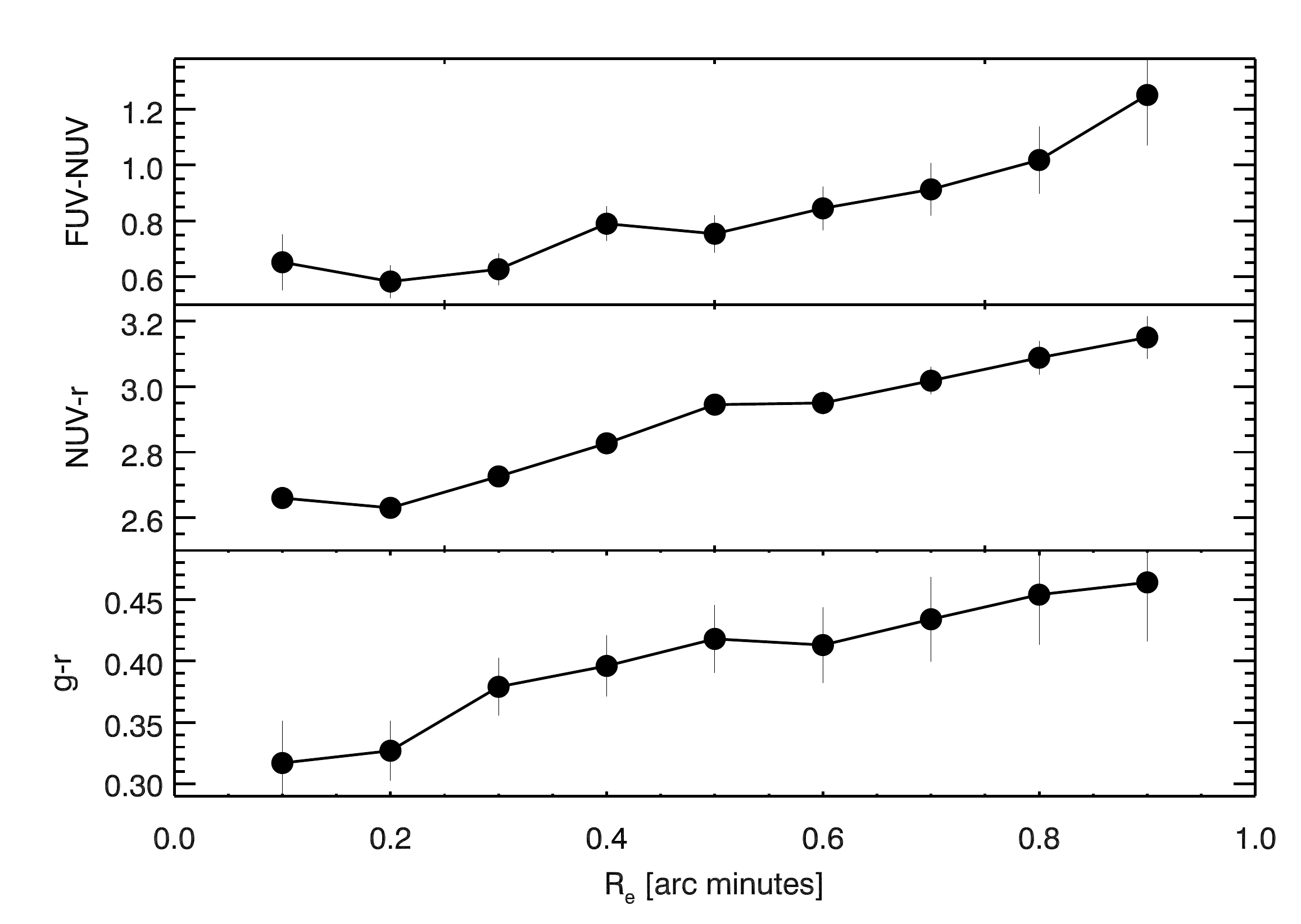}
\caption{\textit{GALEX}+SDSS color profiles of IC~3418's disk using $6\arcsec$ annuli; masked as in Figure~\ref{SED}}\label{profiles}.
\end{center}
\end{figure}

IC~3418's disk has blue optical colors but red UV colors relative to its optical colors and no H$\alpha$, suggesting that star formation was truncated on order 100~Myr ago (Figure~\ref{SED}).  The disk appears truncated in the FUV relative to the NUV (Figure~\ref{UV}) and shows UV and optical color gradients (Figure~\ref{profiles}) consistent with star formation having ceased first in the outer disk.  Though irregular, the disk shows no signs of tidal disturbance (no structured streams or arms of disk stars).  IC~3418 is not detected in H~\textsc{i} ($M_{\rm H~\textsc{i}}<10^{6.9}~{\rm M}_{\odot}$), though a diffuse component in the tail has not been excluded~\citep{Chung09}.  If IC~3418 is on its first orbit through Virgo, as many cluster members appear to be, it was likely accreted on order 100~Myr ago; 500~Myr being a generic cluster infall time. Together, these observations suggest IC~3418 was recently RP stripped by Virgo's intracluster medium (ICM).

\section{Star-Forming Tail}\label{knots}

\begin{deluxetable*}{lrrrrrrrrc} 
\label{table}
\tablecolumns{9} 
\tablewidth{0pc} 
\tablecaption{Properties of the knots and diffuse regions (see Section~\ref{knots} and Figure~\ref{UV}). Columns 3 and 4 give the distance from IC~3418.  The last column indicates whether an H$\alpha$ knot coincides with the object.} 
\tablehead{ 
\colhead{Name} & \colhead{R.A.}   & \colhead{Decl.}    & \colhead{Dist ($\arcmin$)} &  \colhead{Dist (kpc)}
 & \colhead{FUV}    & \colhead{NUV}  & \colhead{FUV-NUV}  & \colhead{H$\alpha$}}
\startdata 
IC~3418 & 12:29:43.82 &  11:24:09.30 & ... & ... & $17.73\pm0.06$ & $16.88\pm0.03$ & $0.85\pm0.07$ & N\\
 \cline{1-9}
k1*  & 12:29:46.88   & 11:23:04.86   &   1.32  &    6.3  & $21.75\pm0.12$ & $21.83\pm0.03$ & $-0.08\pm0.12$ &  N  \\
k2   & 12:29:47.59   & 11:23:33.36   &   1.12  &    5.4  & $20.36\pm0.06$ & $20.35\pm0.02$ & $ 0.01\pm0.07$ &  N  \\
k3   & 12:29:49.84   & 11:23:06.36   &   1.84  &    8.8  & $20.74\pm0.08$ & $20.81\pm0.02$ & $-0.07\pm0.08$ &  N  \\
k4   & 12:29:50.75   & 11:22:36.35   &   2.33  &   11.2  & $20.88\pm0.08$ & $20.82\pm0.02$ & $ 0.06\pm0.08$ &  Y  \\
k5   & 12:29:52.59   & 11:22:49.85   &   2.56  &   12.3  & $20.12\pm0.06$ & $20.15\pm0.01$ & $-0.02\pm0.06$ &  Y  \\
k6   & 12:29:52.29   & 11:22:04.86   &   2.96  &   14.2  & $20.76\pm0.08$ & $20.78\pm0.02$ & $-0.01\pm0.08$ &  Y  \\
k7   & 12:29:52.90   & 11:21:54.35   &   3.20  &   15.3  & $23.4\pm0.3$ & $23.71\pm0.08$ & $-0.3\pm0.3$ &  Y  \\
k8   & 12:29:53.92   & 11:22:01.85   &   3.30  &   15.8  & $21.5\pm0.1$ & $21.48\pm0.03$ & $-0.0\pm0.1$ &  Y  \\
k9   & 12:29:54.53   & 11:21:46.85   &   3.58  &   17.2  & $23.1\pm0.2$ & $23.10\pm0.06$ & $-0.0\pm0.2$ &  Y  \\
k1 ... k9 & ... & ... & ... & ... & $18.60\pm0.03$ & $18.621\pm0.007$ & $-0.02\pm0.03$ & ...\\
 \cline{1-9}
d1*  & 12:29:46.57   & 11:23:10.86   &   1.19  &    5.7  & $20.84\pm0.09$ & $20.43\pm0.02$ & $ 0.41\pm0.09$ &  N  \\
d2   & 12:29:50.65   & 11:22:52.85   &   2.13  &   10.2  & $21.89\pm0.13$ & $21.65\pm0.03$ & $ 0.24\pm0.14$ &  Y  \\
d3   & 12:29:52.79   & 11:22:27.35   &   2.82  &   13.5  & $22.02\pm0.14$ & $21.73\pm0.03$ & $ 0.29\pm0.15$ &  N  \\
d1 ... d3 & ... & ... & ... & ... & $20.25\pm0.06$ & $19.90\pm0.014$ & $0.35\pm0.06$ & ...\\
\enddata
\end{deluxetable*}

We visually separate the tail into bright circular knots and diffuse, asymmetric, lower surface brightness regions.  The resolution of \textit{GALEX} limits our interpretation of these objects. To measure luminosities, regions are defined using the outermost closed NUV surface brightness contour around each object, which are labeled in Figure~\ref{UV}.  A discrete set of contours was considered, but given their dense spacing on the sky this should not impact our results.

The bright knots have NUV luminosities ($\nu f_{\nu}$) between $1.3\times10^{5}~L_{\odot}$ and  $3.4\times10^{6}~L_{\odot}$.  The diffuse regions are redder in FUV-NUV than the knots; the average colors are $0.35\pm0.06$ and  $-0.02\pm0.03$ respectively (Table 1). The knots' colors are consistent with ongoing star formation.  The total FUV luminosity of the tail, $3.3\times10^7~L_{\odot}$, corresponds to a star formation rate (SFR) of $6\times10^{-3}~\rm{ M_{\odot}~yr^{-1}}$ if star formation is ongoing~\citep{Salim07}. This is a lower limit; the knots may contain dust. A knot is also defined (k1) using the innermost FUV contour within the streamer at the tail's base (d1).  Its FUV-NUV color is $-0.08\pm0.12$, typical of the knots. H$\alpha$ is found in the majority of the knots, indicating that star formation is ongoing. H$\alpha$ is not observed in the three UV knots closest to IC~3418.  They may be detected in deeper H$\alpha$ imaging if they have a top-light initial mass function, could be a signature of the tail's formation mechanism, or could indicate RPS of gas from the stellar knots themselves. Only one H$\alpha$ region is observed away from a UV peak. 

\section{Environment of IC~3418}
\label{env}

Within a cluster, tidal interactions and the ICM wind can strip mass from a galaxy's disk.  We estimate the effects of both on IC~3418, noting that the lack of an appropriate neighbor excludes galaxy-galaxy tidal interactions.  The H~\textsc{i} in IC~3418's disk was extremely susceptible to RPS, as expected for a faint, low surface brightness galaxy orbiting in a massive cluster. In contrast,  IC~3418's probable tidal radius is beyond its stellar disk, ruling out a tidal origin for the tail.  The dense molecular clouds were unlikely to be RP stripped as clouds.

We first estimate the impact of RPS using the stripping condition of~\citet{Gunn72}, 
\begin{equation}
\rho_{\rm ICM} v_{\rm orbital}^2 > 2\pi G\sigma_{\ast}\sigma_{\rm H}.
\label{rps}
\end{equation}
We use the usual $\beta$ profile for the ICM density and adopt values of the profile slope, $\beta$, and the central density, $\rho_0$, for Virgo from~\citet{Sanderson03}.  Assuming a three-dimensional distance from the cluster's core of $\sqrt{2}r_{\rm projected}$ and an orbital velocity of $v_{\rm orbital}=1000~\rm{km~s^{-1}}$, the ram pressure is $P_{\rm ram}=\rho_{\rm ICM} v_{\rm orbital}^2\approx40~\rm{M_{\odot}~pc^{-3}~km^2~s^{-2}}$. 

To estimate the restoring pressure on the gas disk,  $P_{\rm rest}=2\pi G\sigma_{\ast}\sigma_{\rm H}$, we first determine the stellar surface density, $\sigma_{\ast}$, from \textit{H}-band observations.  We next estimate the initial gas surface density, $\sigma_{\rm H}$, by assuming a pre-stripping H~\textsc{i} mass and disk size based on IC~3418's morphology, optical luminosity, and optical radius.  The \textit{H}-band magnitude and effective surface brightness of IC~3418 are $m_{\rm H} = 12.72$ and $\mu_e=22.24~{\rm mag~arcsec}^{-2}$~\citep{Gavazzi00, Gavazzi03}.  Assuming a mass to light ratio of 0.78~\citep{Kirby08}, $\sigma_{\ast} = 12~\rm{M_{\odot}~pc^{-2}}$ and $M_{\ast} = 10^{8.9}~{\rm M}_{\odot}$. \citet{Gavazzi05} report an H~\textsc{i} deficiency (see \citet{Giovanelli83}) and an upper limit on the  H~\textsc{i} mass that correspond to an expected $10^{8.8}~{\rm M}_{\odot}$ of H~\textsc{i} in IC~3418. We estimate that IC~3418 contained 70\% of its H~\textsc{i}  within $1\arcmin.85$ by converting its UGC optical radius to a Holmberg radius~\citep {Dickel78} and taking an average ratio of the H~\textsc{i} to optical disk size from~\citet{Giovanelli83}. This gives $\sigma_{\rm H}\approx2~\rm{M_{\odot}~pc^{-2}}$ and $P_{\rm rest} \approx 6~\rm{M_{\odot}~pc^{-3}~km^2~s^{-2}}$.  A large initial gas fraction is consistent with IC~3418's blue optical-H color and low surface brightness. 

We also considered the contribution of a  Navarro-Frenk-White (NFW) dark matter halo to $P_{\rm rest}$; $P_{\rm rest,DM}=\sigma_{\rm H}\left(\partial\phi_{\rm NFW}/\partial r\right)$.  We found that the dark halo makes a significant contribution to the restoring pressure, but that  $P_{\rm ram}>P_{\rm rest}$ unless the halo mass exceeds $\approx10^{13}~{\rm M}_{\odot}$.  This is unrealistically high, and IC~3418's H~\textsc{i} disk was extremely susceptible to RPS.

For the case of molecular cloud RPS, we assume spherically symmetric clouds with conservative values for a molecular cloud size and density of $r_{\rm MC} = 50$~pc and $\rho_{\rm MC} = 10^3~{\rm cm}^{-3}$. This gives a surface density, $\sigma_{\rm H_2}=(4/3)\rho_{\rm MC}r_{\rm MC}$, of $\approx 2000~\rm{M_{\odot}~pc^{-2}}$.  The restoring force from the stellar disk is $P_{\rm rest}=2\pi G\sigma_{\ast}\sigma_{\rm H_2}\approx6000~\rm{M_{\odot}~ pc^{-3}~ km^2~s^{-2}}$.  For molecular clouds, $P_{\rm rest}\gg P_{\rm ram}$.

If IC~3418 is on a purely radial orbit and on its first pass through the cluster potential, then its tidal radius can be estimated by setting
\begin{equation}
\frac{\partial}{\partial r}\frac{Gm_{\rm IC~3418}(r_t)}{r_t} = -r_t\frac{\partial}{\partial r}\frac{GM_{\rm Virgo}(r_o)}{r_o^2}
\label{tidal1}
\end{equation}
and assuming NFW potentials. Here $m(r)$ denotes the mass enclosed within $r$, $r_t$ the galaxy's tidal radius, and $r_o$ the orbital radius within Virgo. This gives an upper limit on $r_t$; a more circular orbit will have a smaller tidal radius at IC~3418's current position.  A lower limit can be estimated by using  
\begin{equation}
\frac{m_{\rm IC~3418}(r_t)}{m_{\rm IC~3418}} = 0.58\left(\frac{r_o}{r_{v,\rm{Virgo}}}\right)^{0.50}
\label{tidal2}
\end{equation}
 \citep{Mamon00}, which assumes that the galaxy is at pericenter and that it has an orbital eccentricity typical of subhalos in \textit{N}-body simulations. Here $r_{v,{\rm Virgo}}$ is Virgo's virial radius. For both estimates we assume a dark matter to baryon mass ratio of $M_{\rm DM}/M_{\rm bary}\approx15$ ($M_{\rm DM}\approx10^{10.3}~{\rm M}_{\odot}$). We adopt $r_{v,\rm{Virgo}}$ from \citet{Mclaughlin99}. 
 
IC~3418's current tidal radius is between 9 and 30~kpc. If the galaxy is not at pericenter, then its final tidal radius will be within 9~kpc. For comparison, its \textit{H}-band effective radius is 4.6~kpc.  While it is unlikely that IC~3418's disk is being tidally stripped, the tail may be tidally stripped from the galaxy as it passes through Virgo's core, thereby contributing to the ICL.
 
\section{Morpholgy}
\label{morphology}
Tails and arcs of young stars and/or H~\textsc{ii} regions which are consistent with an RPS origin have been observed in several clusters.  C07 discovered two tails of young stars in deep \textit{Hubble Space Telescope} imaging.  One, an $\sim0.1~L^{\ast} $ galaxy well within Abell 1689's X-ray halo, should be completely RP stripped, which is consistent with the declining SFR in the galaxy's disk observed by C07.  The other, an $L^{\ast}$ galaxy in Abell 2667, should be partially stripped~(C07).  Its asymmetric appearance and nuclear burst of star formation match simulations of partially RP stripped disks \citep{Schulz01, Kronberger08}.  Using Equation~(\ref{tidal2}) and the same choice of parameters as C07, we estimate tidal radii beyond the stellar disk for both galaxies. \citet[][Y08]{Yoshida08} discovered a tail of star forming knots and young stellar filaments behind RB 199 in deep $B$, $R_C$, and H$\alpha$ observations of the Coma cluster.  Y08 estimate that the stellar disk is not tidally stripped, but that the gas disk is partially RP stripped. The galaxy ESO 137-001 in Abell 3627, a very dynamically young nearby cluster, was discovered to have an extended X-ray tail in which multiple H~\textsc{ii} regions were detected~\citep{Sun06, Sun07}. 

Galaxies with arcs of H~\textsc{ii} regions following the bow shock morphology of the H~\textsc{i} have been observed in Virgo~\citep{Kenney99} and Abell 1367~\citep{Gavazzi01}.  The morphology of these arcs suggests cloud stripping, but cloud formation likely occurred in the shock compressed gas while clouds in the outer disk were dissociated and then stripped. Isolated dense clouds can be ablated by the ICM wind~\citep{Nulsen82, Raga98}, dissociated if a shock is driven directly into them~\citep{Aannestad73}, and heated by star formation and then stripped~\citep{Kenney04}.

\begin{figure}
\begin{center}
\includegraphics[height=1.8in]{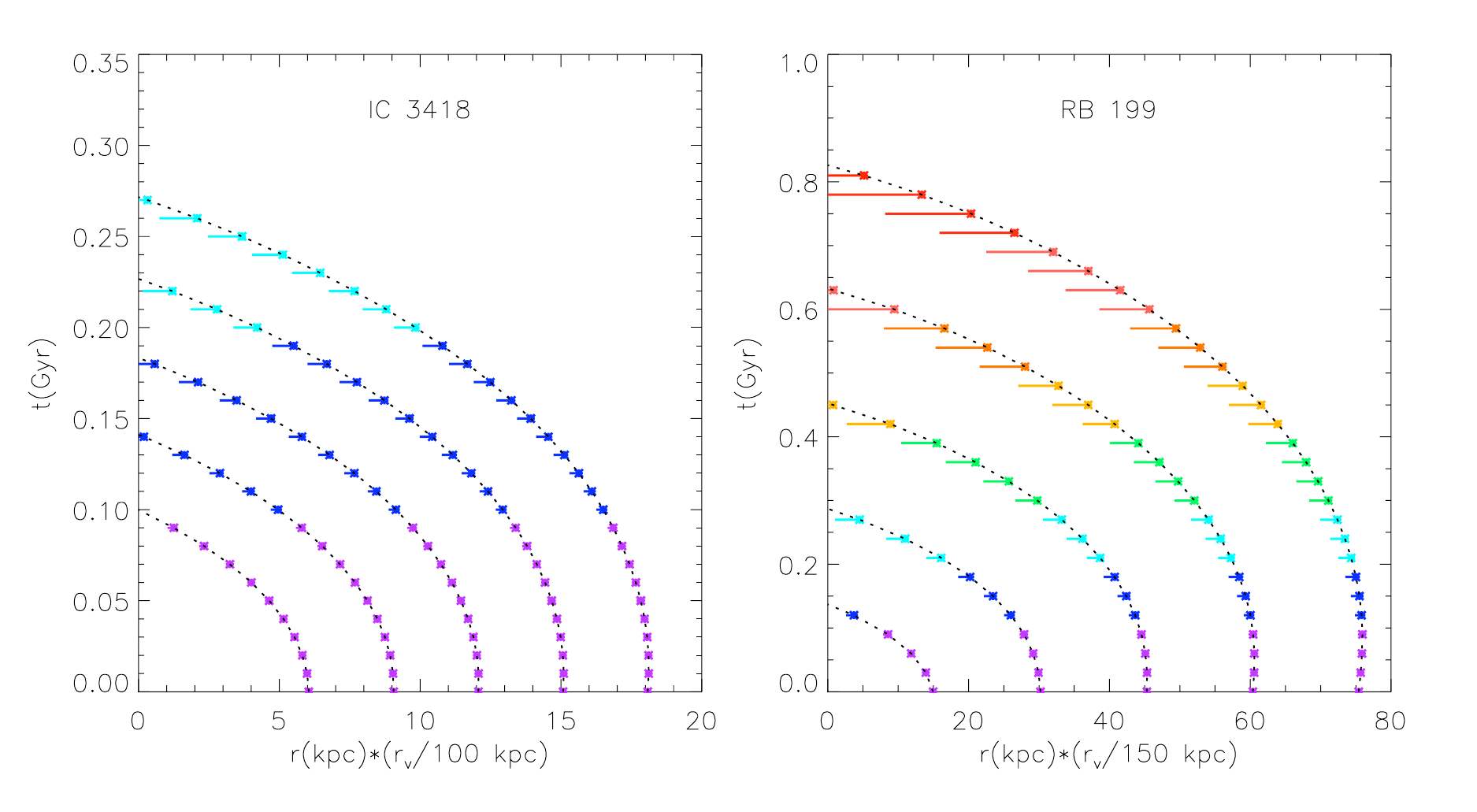}
\caption{Simple model of tail morphology for tails resembling IC~3418 (left) and RB~199 (right). Tracks represent a filament falling back onto the galaxy.  Line segments show the diverging orbits of two episodes of star formation separated by $t_{\rm form}=100$~Myr.  The divergence of the orbits is due to an additional $a_{\rm MC}$ which is applied to the later forming stars for the first $t_{\rm form}$.  The segments change color every 100~Myr.}
\label{model}
\end{center}
\end{figure}

We advance a basic mechanism to account for the tails' morphologies, which are characterized by isolated knots, long filaments, and intermediate ``streamers" such as that in IC~3418 (d1/k1).  Giant molecular clouds may spawn multiple episodes of star formation. The cloud experiences a force from the ICM wind, but the newly formed stars do not.  Due to the differing accelerations, the newly formed stars are essentially dropped from the cloud, which itself is supported by the ICM wind. The cloud trails a filament of stars.

The details of the tails' morphologies provide a probe of their formation.  To first order, a cloud forms a filament of length $(1/2)a_{\rm MC}\Delta t_{\rm form}^2$, where $a_{\rm MC}$ is the acceleration due to the ICM wind, $a_{\rm ICM}$, on the molecular cloud and $\Delta t_{\rm form}$ is the time between the cloud's first and last episodes of star formation.  In the cloud's frame, the stars accelerate toward the galaxy with $a_{\rm MC}$.  Large $a_{\rm MC}$ create long filaments.  Accordingly, the long stellar filaments behind RB~199 are indicative of large $a_{\rm MC}$, itself characteristic of dynamical coupling between the gas phases and therefore of cloud stripping.  The $a_{\rm ICM}$ can be estimated as $P_{\rm ram}/\sigma_{\rm gas}$, or, for the H~\textsc{i} disk,
\begin{displaymath}
a_{\rm H~\textsc{i}}=0.02~\rm{kpc~Myr^{-2}}\left(\frac{P_{\rm ram}}{40~\rm{M}_{\odot}~pc^{-3}~km^2~s^{-2}}\right)
\end{displaymath}
\begin{displaymath}
\times\left(\frac{2~{\rm M}_{\odot}~{\rm pc}^{-2}}{\sigma_{\rm H~\textsc{i}}}\right),
\end{displaymath}
which can form a 10 kpc filament in 30 Myr. Assuming molecular clouds that are not dynamically coupled to the H~\textsc{i}, $a_{\rm H_2}/a_{\rm H}=\sigma_{\rm H_2}/\sigma_{\rm H} \approx 10^{-3}$ (see Section~\ref{env}), corresponding to much shorter ``filaments". 

The treatment above assumes that the cloud and its filament orbit together.  In contrast, both tides, which increase the distance between two episodes as they fall, and the ICM wind, which continuously increases the orbital energy of a cloud while it exists, work to lengthen filaments.  Therefore, if $a_{\rm MC}=a_{\rm H}$, then \textit{only} long filaments are formed, which conflicts with the observations. We show below that a plethora of morphologies can be formed with $a_{\rm MC}=a_{\rm H_2}$.

The influences of both gas physics and gravity are observed, requiring us to consider both forces. The streamer in IC~3418's tail and the filaments observed by Y08 and C07 are located close to the galaxy, where tidal forces are strongest.  The extended objects are also redder, indicating that they are older and therefore potentially more dynamically evolved. On the other hand, the streamer in IC~3418 shows a strong UV color gradient and, in RB~199, when a star forming knot is associated with a filament, it is found downwind.  Both highlight the role of the ICM wind.

Figure~\ref{model} illustrates the next step in modeling complexity. We postulate clouds that are generated with a moderate initial velocity heading away from the galaxy. These clouds fall back toward the galaxy. Each track with its related line segments represents two episodes of star formation separated by $t_{\rm form}=100$~Myr, which is motivated by the UV gradient in the streamer, as they fall in an NFW potential. Both episodes have the same initial orbit, but the higher, later forming, episode has an additional acceleration for the first $t_{\rm form}$.  The line segments represent potential knots and filaments. 

In simulations, RP stripped gas is accelerated to the rest velocity of the ICM, but travels away from the galaxy as it accelerates. Gas located close to the galaxy has a lower velocity with respect to the galaxy and turbulent flow around the disk can cause gas to actually fall back onto the galaxy from the wake \citep[][R08]{Roediger08}. In the model, increasing a cloud's initial velocity increases the filament length until the velocity exceeds the escape velocity and the knot speeds away without forming a filament.

The left and right panels of Figure~\ref{model} represent the tails of IC~3418 and RB~199, respectively. They are differentiated by their overall length scale and age.  In each, $P_{\rm ram}$ is within the range allowed by observations.  In simulations, tail length is determined by the mass loss per orbital length, which will vary considerably between galaxies (R08).

Several suggestive features are reproduced in Figure~\ref{model}.  Knots, short streamers, and long filaments all form.  As seen in the observations, the filaments' lengths and ages are correlated, particularly for neighboring objects, and in each tail the longest filaments are found closest to the galaxy.  For a nice example of the first feature, compare d2 and its surrounding knots. That a simple variation in the length scale reproduces both IC~3418 and RB~199 is an additional success. Tail morphologies are rich with information about their formation mechanisms, which should motivate further modeling.  The current work indicates that the observations favor in situ cloud formation, or at least $a_{\rm MC}=a_{\rm H_2}$.

\section{Discussion}
RPS of H~\textsc{i} is the only theoretically favored means of removing mass from IC~3418's disk at its current position in the Virgo Cluster.  Nevertheless, star formation is occurring in the galaxy's wake, requiring the presence of $\rm{H_2}$.  We suggest that $\rm{H_2}$ forms within the stripped wake aided by turbulence.  Turbulence in the wakes of RP stripped galaxies has been seen in both simulations (R08) and observations~\citep{Yoshida02}.  In simulations, turbulence can drive both the formation of star clusters within clouds~\citep{Padoan02} and cloud formation itself~\citep{Maclow04}.  The quasi-scale free hierarchical nature of turbulence works to enhance density contrasts in a turbulent region, increasing the chance that gas can cool and collapse.  An alternate theory requires that the $\rm H_2$ in the disk is tightly dynamically bound to the H~\textsc{i}, though no mechanism for this binding is advanced.  We have shown that the morphologies of the tails disfavor this interpretation.

Further constraining the tails' formation mechanism, and in particular the possible role of turbulence, will require additional observations and improved modeling, presumably using hydrodynamical simulations. Deep, high resolution, imaging of one or more tails in multiple optical bands, UV, and H$\alpha$, for which IC~3418 is a prime candidate, could provide detailed constraints on a model of tail formation. 

IC~3418's tail represents a novel environment in which to study cloud and star formation. The SFR within the normal star forming disk of a massive galaxy correlates with the gas surface density~\citep{Toomre64, Kennicutt89, Kennicutt98, Martin01}.  In other environments, the SFR does not follow a Schmidt-Kennicutt Law. Examples are extended gas disks~\citep{Ferguson98, Thilker07, Bigiel08}, low surface brightness disks~\citep{Wyder09}, and dwarf irregular galaxies~\citep{Hunter06, Lee09}.   Gravitational effects,  turbulence, magnetic fields, and thermal support may all play a role in cloud and star formation. Determining the role that each plays in driving and regulating star formation will require studying multiple regimes. \\
\\
This research has made use of the GOLD Mine Database and of the NASA/IPAC Extragalactic Database (NED) which is operated by the Jet Propulsion Laboratory, California Institute of Technology, under contract with the National Aeronautics and Space Administration. The authors thank L. Cortese for helpful comments.

\end{document}